\input{epsfig}
\documentstyle[prd,eqsecnum,aps,twocolumn]{revtex}
\def\lambdabar{\hbox{$\lambda$\kern-.53em\raise.2em\hbox{--}}}

\catcode`\@=11

\def\maketitle2{\par 
\begingroup
\let\cite\@bylinecite
\def\thefootnote{\fnsymbol{footnote}}%
\twocolumn[\@maketitle2\vskip2pc]%
\thispagestyle{plain}\@thanks
\endgroup
\def\thefootnote{\arabic{footnote}}%
\setcounter{footnote}{0}%
\let\maketitle2\relax \let\@maketitle2\relax
\let\@thanks\relax \let\@authoraddress\relax \let\@title\relax
\let\@date\relax \let\thanks\relax \let\@abstract\relax 
\let\@pacs\relax}

\def\abstract#1{\gdef\@abstract{{\par 
\bgroup
\ifdim\prevdepth=-1000pt \prevdepth0pt\fi
\hsize\columnwidth
\dimen0=-\prevdepth \advance\dimen0 by17.5pt \nointerlineskip
\small\vrule width 0pt height\dimen0 \relax}{~~}#1\egroup}}

\def\pacs#1{\gdef\@pacs{{\par 
\bgroup
\hsize\columnwidth \parindent0pt
\ifdim\prevdepth=-1000pt \prevdepth0pt\fi
\dimen0=-\prevdepth \advance\dimen0 by20pt\nointerlineskip
\egroup} PACS numbers:~#1}}

\def\@maketitle2{
\@preprint
\@title
\ifdim\prevdepth=-1000pt \prevdepth0pt\fi
\@authoraddress
\@date
\begin{list}{}{\leftmargin=0.10753\textwidth \rightmargin=\leftmargin
\itemsep=1pc\partopsep=-1pc}
\item\@abstract
\item\@pacs
\end{list}
}

\catcode`\@=12

\begin{document}
\draft
\preprint{LA-UR-98-2347}
\title{How Wigner Functions Transform Under Symplectic Maps}
\author{Alex J. Dragt$^{1,*}$ and Salman Habib$^{2,\dagger}$}
\address{$^1$Center for Theoretical Physics, University of Maryland,
College Park, MD 20742} 
\address{$^2$T-8, Theoretical Division, MS B285, Los Alamos National
Laboratory, Los Alamos, NM 87545}
\date{\today}

\abstract
{It is shown that, while Wigner and Liouville functions transform in
an identical way under linear symplectic maps, in general they do not
transform identically for nonlinear symplectic maps.  Instead there
are ``quantum corrections'' whose $\hbar \rightarrow 0$ limit may be
very complicated.  Examples of the behavior of Wigner functions in the
$\hbar \rightarrow 0$ limit are given in order to examine to what
extent the corresponding Liouville densities are recovered.}

\pacs{}

\maketitle2
\narrowtext

\section{Introduction}
\label{sec:level1}
Consider an ensemble of non-interacting particles.  Classically such
an ensemble may be described by a Liouville density in phase space.
Denote the collection of phase-space variables by the symbol $z$,
\begin{equation}
z = (q_1,q_2, q_3; p_1, p_2, p_3).
\end{equation}
Let $w(z)$ be the Liouville density at the point $z$.  Then, by
definition, the number of particles $d^6 N$ in the phase-space volume
$d^6z$ is given by the relation
\begin{equation}
d^6 N = w(z) d^6 z.
\end{equation}

Suppose the ensemble of particles is {\em initially} characterized by
a distribution function $w^i$.  Let ${\cal M}$ be a symplectic map
(canonical transformation) that sends {\em initial} points $z^i$ to
{\em final} points $z^f$,
\begin{equation}
z^f = {\cal M} z^i.
\end{equation}
This map has the associated Jacobian matrix $M(z^i)$ defined by the
relation
\begin{equation}
M_{ab}(z^i) = \partial z^f_a/\partial z^i_b.
\end{equation}
By the definition of ${\cal M}$ being a symplectic map, the matrix $M$
must be a symplectic matrix,
\begin{equation}
M^TJM = J.
\end{equation}
Here $J$ is the usual fundamental matrix having the $3 \times 3$ block
form 
\begin{equation}
J = \left( \begin{array}{cc} 0 & I \\
-I & 0\end{array} \right).
\end{equation}
It can be shown from (1.5) that $M$ has unit determinant \cite{1},
\begin{equation}
\det M = 1.
\end{equation}
Correspondingly, if $w^f$ is the {\em final} distribution function
resulting from the action of ${\cal M}$, Liouville's theorem follows
as a consequence of (1.7) and (1.2),
\begin{equation}
w^f(z^f) = w^i(z^i).
\end{equation}
Now insert (1.3) into (1.8) to find the relation
\begin{equation}
w^f(z^f) = w^i ({\cal M}^{-1} z^f).
\end{equation}
Since $z^f$ is a generic point, the relation (1.9) may equally well be
written in the form
\begin{equation}
w^f(z) = w^i ({\cal M}^{-1} z).
\end{equation}
This equation describes how the Liouville function transforms under
the action of a general symplectic map.

Next consider a {\em quantum} description of this same ensemble of
noninteracting particles.  In the quantum description, an ensemble is
characterized by a density operator $\hat{\rho}$.  Associated with any
density operator is a Wigner function $W(z)$ that in some ways is the
quantum analog of the Liouville function $w(z)$ \cite{2a}.  Also,
there are unitary transformations ${\cal U}$ that are the quantum
analog of the symplectic maps ${\cal M}$.  It is therefore tempting to
consider the possibility
\begin{equation}
W^f(z) \stackrel{?}{=} W^i ({\cal M}^{-1} z).
\end{equation}
It is known that (1.11) is true in the case that ${\cal M}$ is a {\em
translation} or a {\em linear} transformation \cite{2}.  The purpose
of this note is to show, by a simple counter-example, that (1.11) is
not in general true for a nonlinear symplectic map.  Section 1.2
describes needed Lie-algebraic tools in the classical case.  Section
1.3 describes quantum concepts and defines the Wigner function.
Section 1.4 presents the counter-example, and discusses the $\hbar
\rightarrow 0$ limit for sample Wigner functions.  A final section
discusses various aspects of what is already known and available in
the literature about Wigner functions, and shows that in this context
the failure of (1.11) in the nonlinear case is no surprise.

Why should transformation properties of the Wigner function be of
interest? 

First, there is the general question of the relation between quantum
and classical mechanics and the way that quantum mechanics becomes
classical in the limit $\hbar \rightarrow 0$.  Second, a light beam
may be described, using ray theory, in terms of a Liouville function.
In wave theory, where the ``reduced'' wavelength $\lambdabar$ plays a
role analogous to Planck's constant, the same beam may be described by
a Wigner function.  In the ray description, aberration effects are
characterized by nonlinear symplectic maps \cite{3}.  Correspondingly,
in a wave description, one would like to know the effect of
aberrations on the Wigner function.

\section{Lie Algebraic Tools}
\label{sec:level2}
Let $f(z)$ be any function of the phase-space variables $z$.
Associated with $f$ is a {\em differential} operator, called a {\em
Lie} operator, that will be denoted by the symbol $:\!\!f\!\!:$ and is
defined by the rule
\begin{equation}
:\!\!f\!\!: = \sum_i (\partial f/\partial q_i)(\partial /\partial
p_i) - (\partial f/\partial p_i)(\partial /\partial q_i).
\end{equation}
When $:\!\!f\!\!:$ acts on any other phase-space function $g(z)$, the
result is a Poisson bracket,
\begin{equation}
:\!\!f\!\!:g = [f,g].
\end{equation}

Powers of Lie operators are defined by repeated application, with
$:\!\!f\!\!:^0$ defined to be the identity operator,
\begin{eqnarray}
:\!\!f\!\!:^0 g &=& g, \nonumber \\
:\!\!f\!\!:^1 g &=& [f,g] \nonumber \\
:\!\!f\!\!:^2 g &=& :\!\!f\!\!:(:\!\!f\!\!:g) = [f,[f,g]], \ {\rm
etc.}
\end{eqnarray}
Once powers have been defined, power series are also defined.  Of
particular interest is the exponential series,
\begin{equation}
\exp (:\!\!f\!\!:) = e^{:\!f\!:} = \sum_{\ell = 0}^{\infty}
:\!\!f\!\!:^{\ell}/\ell !.
\end{equation}
This series is called a {\em Lie transformation}. According to (2.3)
and (2.4), a Lie transformation acts on the function $g(z)$ to give
the result
\begin{equation}
\exp (:\!\!f\!\!:) g = g + [f,g] + [f,[f,g]]/2! + \cdots .
\end{equation}

Suppose $f$ is any function of the initial phase-space variables
$z^i$.  Define final variables $z^f$ by the rule
\begin{equation}
z^f_a = \exp (:\!\!f(z^i)\!\!:) z^i_a.
\end{equation}
Comparison of (1.3) and (2.6) indicates that in this case the map
${\cal M}$ can be written in the form
\begin{equation}
{\cal M} = \exp (:\!\!f\!\!:).
\end{equation}
It can be shown that any map of the form (2.7) is a symplectic map.
Conversely, any symplectic map can be written as a product of Lie
transformations \cite{1}.

Three simple Lie transformations are of particular interest for
subsequent use.  All apply to the case of a 2-dimensional phase space
described by the variables $q$ and $p$.  First consider the symplectic
map given by (2.7) when $f$ has the form
\begin{equation}
f(z) = \alpha q.
\end{equation}
Here $\alpha$ is a parameter.  Then one finds from (2.3) and (2.4) the
results
\begin{eqnarray}
q^f &=& {\cal M} q^i = \exp [:\!\!\alpha q^i\!\!:]q^i = q^i, \nonumber
\\ 
p^f &=& {\cal M} p^i = \exp [:\!\!\alpha q^i\!\!:]p^i = p^i + \alpha . 
\end{eqnarray}
Second, suppose $f$ is of the form
\begin{equation}
f(z) = \alpha q^2/2.
\end{equation}
Then one finds the results
\begin{eqnarray}
q^f &=& {\cal M} q^i = \exp [:\!\!\alpha (q^i)^2/2\!\!:]q^i = q^i,
\nonumber \\  
p^f &=& {\cal M} p^i = \exp [:\!\!\alpha (q^i)^2/2\!\!:]p^i = p^i +
\alpha q^i. 
\end{eqnarray}
Finally, suppose $f$ is of the form
\begin{equation}
f(z) = \alpha q^3/3.
\end{equation}
Then one finds the results
\begin{eqnarray}
q^f &=& {\cal M} q^i = \exp [:\!\!\alpha (q^i)^3/3\!\!:]q^i = q^i,
\nonumber \\ 
p^f &=& {\cal M} p^i = \exp [:\!\!\alpha (q^i)^3/3\!\!:]p^i = p^i +
\alpha (q^i)^2. 
\end{eqnarray}
Transformation (2.9) is a translation, (2.11) is a linear
transformation, and (2.13) is a nonlinear transformation.

\section{Quantum Concepts}
\label{sec:level3}
For simplicity of presentation, the quantum treatment will be
described for the case of a 2-dimensional phase space and the
associated operators $Q$ and $P$.  Generalization to higher dimensions
will be obvious.

Suppose $\hat{A}$ is any operator (presumably involving $Q$ and $P$).
Associated with this operator is the function $A(q,p)$, called the
{\em Weyl transform} of $\hat{A}$, and defined by the rule
\begin{equation}
A(q,p) = \int dy \langle q + y/2 |\hat{A}| q - y/2\rangle \exp (-i py/\hbar
).
\end{equation}
Here $|q\rangle$, as usual, denotes a position eigenstate of the operator
$Q$. 
Conversely, the underlying operator $\hat{A}$ can be recovered from a
knowledge
of $A$ by using an inverse Weyl transformation,
\begin{eqnarray}
\hat{A} (Q,P) &=& (2\pi \hbar )^{-2} \int d\phi d\tau dq dp A(q,p)
\nonumber\\ 
&&\times\exp\{(i/\hbar)[\phi (Q-q) + \tau (P - p)]\}.
\end{eqnarray}
The reader should verify that if
\begin{equation}
\hat{A} = T(P) + V(Q),
\end{equation}
then, by the Weyl correspondence,
\begin{equation}
A = T(p) + V(q),
\end{equation}
and vice versa.  Finally, the Weyl transform has the property that, if
$\hat{A}$ and $\hat{B}$ are any two operators, there is the relation
\begin{equation}
(2\pi \hbar )^{-1} \int dq dp A(q,p) B(q,p) = tr (\hat{A}\hat{B}).
\end{equation}

Consider a quantum ensemble consisting of $N$ non-interacting members.
Let the $j$'th member of the ensemble be described by the state vector
$|\psi_j\rangle$.  Then the density operator $\hat{\rho}$ for this
ensemble is defined by the rule
\begin{equation}
\hat{\rho} = (1/N) \sum^N_{j=1} |\psi_j\rangle \langle \psi_j|.
\end{equation}
Evidently by construction the density operator is Hermitian and positive
definite, and has unit trace.  If $\hat{A}$ is any operator, its
expectation
value for the $j$'th member in the ensemble is $\langle \psi_j
|\hat{A}|\psi_j\rangle$, and the {\em ensemble average} of this expectation
value is
\begin{equation}
[\langle \psi_j |\hat{A}| \psi_j\rangle ]_{ea} = (1/N) \sum^N_{j=1} \langle
\psi_j|\hat{A}|\psi_j\rangle .
\end{equation}
Let $|\chi_n\rangle$ be any orthonormal set of basis vectors.  The
respresentation of the identity $I$ in this basis is given by the relation
\begin{equation}
I = \sum_n |\chi_n\rangle \langle \chi_n|.
\end{equation}
Now insert (3.8) into (3.7) to find the familiar formula that relates a
combined
quantum-mechanical and ensemble average to a trace,
\begin{eqnarray}
[\langle \psi_j |\hat{A}| \psi_j\rangle ]_{ea} &=& (1/N) \sum^N_{j=1}
\langle
\psi_j|\hat{A}|\psi_j\rangle \nonumber \\
&=& (1/N) \sum^N_{j=1} \langle \psi_j|\hat{A}I|\psi_j\rangle
\nonumber\\
&=& (1/N) \sum_n
\sum^N_{j=1} \langle \psi_j|\hat{A}|\chi_n\rangle \langle
\chi_n|\psi_j\rangle
\nonumber \\
&=& \sum_n \langle \chi_n|(1/N) \sum_j|\psi_j\rangle \langle
\psi_j|\hat{A}|\chi_n\rangle \nonumber \\
&=& \sum_n \langle \chi_n|\hat{\rho} \hat{A}|\chi_n\rangle =
tr(\hat{\rho}\hat{A}).
\end{eqnarray}

The Wigner function associated with any density operator $\hat{\rho}$ is
defined
to be the Weyl transform of the operator $\hat{\rho}/(2\pi \hbar )$,
\begin{equation}
W(q,p) = (2\pi \hbar )^{-1} \int dy \langle q + y/2 |\hat{\rho}|
q-y/2\rangle
\exp (-i py/\hbar ).
\end{equation}
It follows from this definition and (3.5) that $W$ has the properties
\begin{equation}
\int dqdp W(q,p) A(q,p) = tr (\hat{\rho} \hat{A}),
\end{equation}
\begin{equation}
\int dqdp W(q,p) = tr (\hat{\rho} I) = tr (\hat{\rho}) = 1.
\end{equation}
The left side of (3.11) resembles a classical phase-space average, and the
right
side of (3.11) is a quantum-mechanical ensemble average.  Thus, in this
respect,
the Wigner function $W$ resembles a normalized Liouville function.  It can
also
be verified that
\begin{equation}
\int dp W(q,p) = \langle q|\hat{\rho}|q\rangle \geq 0,
\end{equation}
\begin{equation}
\int dq W(q,p) = \langle p|\hat{\rho}|p\rangle \geq 0,
\end{equation}
so that any partially integrated Wigner function behaves like a position or
momentum density.  However, unlike a Liouville density, in general the
Wigner
function itself may sometimes be {\em negative} \cite{4}.

\section{Simple Counter-Example}
\label{sec:level4}
Suppose that each member of a quantum ensemble is acted upon by a
common unitary transformation ${\cal U}$.  Then the state vector
$|\psi_j \rangle$ for the $j$'th member of the ensemble becomes
$|\psi_j^{\prime}\rangle$ with
\begin{equation}
|\psi_j^{\prime}\rangle = {\cal U} |\psi_j\rangle,
\end{equation}
and correspondingly, according to (3.6), the density operator
$\hat{\rho}$ for the ensemble becomes $\hat{\rho}^{\prime}$ with
\begin{equation}
\hat{\rho}^{\prime} = {\cal U}\hat{\rho} \ {\cal U}^{-1}.
\end{equation}
It follows from the definition (3.10) that the transformed Wigner
function $W^{\prime}$ associated with $\hat{\rho}^{\prime}$ is given
by the relation
\begin{eqnarray}
W^{\prime} (q,p) &=& (2\pi \hbar)^{-1} \int dy \langle q + y/2|{\cal
U}\hat{\rho}
\ {\cal U}^{-1}|q-y/2\rangle \nonumber\\
&&\times \exp (-ipy/\hbar ).
\end{eqnarray}

In analogy with (2.7) and (2.8), suppose ${\cal U}$ is of the form
\begin{equation}
{\cal U} = \exp (i\alpha Q/\hbar ).
\end{equation}
(This analogy is motivated by the close resemblance between the
Poisson bracket Lie algebra of Classical Mechanics and the commutator
Lie algebra of Quantum Mechanics.)  For a position eigenstate
$|q\rangle$ there is the result
\begin{equation}
{\cal U}|q\rangle = \exp (i\alpha Q/\hbar )|q\rangle = \exp (i\alpha
q/\hbar )
|q\rangle,
\end{equation}
and for the momentum eigenstate $|p\rangle$ there is the result
\begin{equation}
{\cal U}|p\rangle = \exp (i\alpha Q/\hbar )|p\rangle = |p + \alpha
\rangle. 
\end{equation}
This latter result follows from the fundamental relation
\begin{equation}
|p \rangle = (2\pi \hbar )^{-1/2} \int dq \exp (ipq/\hbar ) |q\rangle, 
\end{equation}
or, equivalently, from the fundamental commutation rule
\begin{equation}
\{ Q,P\} = QP - PQ = i\hbar I.
\end{equation}
Evidently, (4.5) and (4.6) are the quantum analog of (2.9).

What is $W^{\prime}$ when ${\cal U}$ is given by (4.4)?  Examine the
ingredients of the integrand for (4.3). From (4.5) it follows that
\begin{equation}
{\cal U}^{-1} |q-y/2\rangle = \exp [-i\alpha (q-y/2)/\hbar ]
|q-y/2\rangle, 
\end{equation}
\begin{equation}
\langle q+y/2| {\cal U} = \exp [+i\alpha (q+y/2)/\hbar ]\langle q +
y/2|. 
\end{equation}
Hence, (4.3) becomes
\begin{eqnarray}
W^{\prime} (q,p) = &&(2\pi \hbar )^{-1} \int dy \langle
q+y/2|\hat{\rho}|q-y/2\rangle \nonumber \\
&&\times\exp [-i\alpha (q-y/2)/\hbar ] \exp [+i\alpha (q+y/2)/\hbar
]\nonumber\\ 
&&\times \exp(-ipy/\hbar).
\end{eqnarray}
Moreover, addition of the exponents in (4.11) yields 
\begin{eqnarray}
&&[-i\alpha (q-y/2)/\hbar ] + [+i\alpha (q+y/2)/\hbar ] + (-ipy/\hbar
) \nonumber\\
&&=-i(p-\alpha ) y/\hbar .
\end{eqnarray}
Consequently, (4.3) takes the final form
\begin{eqnarray}
W^{\prime} (q,p) &=& (2\pi \hbar )^{-1} \int dy \langle q +
y/2|\hat{\rho}|q-y/2\rangle \nonumber\\
&&\times \exp [-i(p-\alpha ) y/\hbar ]\nonumber\\
&=& W(q,p-\alpha ).
\end{eqnarray}
From (2.9) it follows that
\[
{\cal M}^{-1} q = \exp (-:\!\!\alpha q\!\!:) q = q,
\]
\begin{equation}
{\cal M}^{-1} p = \exp (-:\!\!\alpha q\!\!:) p = p-\alpha .
\end{equation}
Therefore, inspection of (4.13) and (4.14) illustrates that (1.11)
holds for a translation.

Next suppose that ${\cal U}$ is of the form
\begin{equation}
{\cal U} = \exp [i\alpha Q^2/(2\hbar )].
\end{equation}
The calculation proceeds as before.  For example, there is the
relation 
\begin{equation}
{\cal U}^{-1} |q-y/2\rangle = \exp [-i\alpha (q-y/2)^2/(2\hbar )]
|q-y/2\rangle,
\end{equation}
and the ``exponent addition'' relation analogous to (4.12) takes the
form  
\begin{eqnarray}
&&[-i\alpha (q-y/2)^2/(2\hbar )] + [+i\alpha (q+y/2)^2/(2\hbar )] +
(-ipy/\hbar )\nonumber\\
&&= -i (p-\alpha q)y/\hbar .
\end{eqnarray}
The reader is urged to verify this result in the solitude of some
evening.  It follows that in this case
\begin{equation}
W^{\prime} (q,p) = W(q,p-\alpha q),
\end{equation}
and consequently (1.11) again holds.

Finally, suppose that ${\cal U}$ is of the form
\begin{equation}
{\cal U} = \exp [i\alpha Q^3/(3\hbar )].
\end{equation}
In this case the exponent addition relation is
\begin{eqnarray}
&&[-i\alpha (q-y/2)^3/(3\hbar )] + [+i\alpha (q+y/2)^3/(3\hbar )] +
(-ipy/\hbar
) \nonumber \\
&&= -i(p-\alpha q^2)y/\hbar + i\alpha y^3/(12 \hbar ).
\end{eqnarray}
Correspondingly, $W^{\prime}$ is given by the relation
\begin{eqnarray}
W^{\prime} (q,p)&=&(2\pi \hbar )^{-1} \int dy \langle
q+y/2|\hat{\rho}|q-y/2\rangle\nonumber\\
&\times&\exp [-i(p-\alpha q^2)y/\hbar ] \exp [i\alpha y^3/(12\hbar )].
\end{eqnarray}
Evidently (1.11) now {\em fails} to hold due to the offensive term
$\exp [i\alpha y^3/(12 \hbar )]$ in the integrand of (4.21).

At this point we remark that a similar calculation shows that (1.11)
fails for any ${\cal U}$ of the form $\exp [i\alpha Q^n/(n\hbar )]$
with $n \geq 3$.  Moreover, it can be shown that any symplectic map
can be written as a product of linear transformations and
transformations of the form $\exp (\alpha :\!\!q^n\!\!:/n)$ \cite{5a}.
If follows that (1.11) fails for all nonlinear transformations.

What more can be said?  Suppose the offensive term is expanded in a
Taylor series,
\begin{equation}
\exp [i\alpha y^3/(12\hbar )] = 1 + i\alpha y^3/(12\hbar ) + \cdots.
\end{equation}
Insertion of this expansion into the integral (4.21) gives the result
\begin{eqnarray}
W^{\prime}(q,p) &=& W(q,p-\alpha q^2) \nonumber \\
&&+ i\alpha /[(2\pi \hbar )(12\hbar )] \int dy \langle q
+
y/2|\hat{\rho}|q-y/2\rangle \nonumber\\
&&\times\exp [-i(p-\alpha q^2) y/\hbar ] \left[y^3 + \cdots \right]. 
\end{eqnarray}
Powers of $y$ in the integrand can be obtained by differentiating the
factor $\exp (-ipy/\hbar )$ with respect to $p$.  Consequently, (4.23)
can be recast in the form
\begin{eqnarray}
&&W^{\prime}(q,p) \nonumber\\
&&= W(q,p-\alpha q^2) + \alpha (\hbar^2/12)(\partial
/\partial
p)^3 W(q,p-\alpha q^2) + \cdots .\nonumber\\
\end{eqnarray}

It is tempting to view (4.24) as a quantum correction to (1.11), and
to speculate that (1.11) becomes exact in the limit $\hbar \rightarrow
0$.  This may be true in some cases providing $W$ itself is well
behaved as $\hbar\rightarrow 0$ and remains sufficiently smooth so
that the derivatives occurring in (4.24) remain bounded.  For example,
consider the simple harmonic oscillator described by the quantum
Hamiltonian
\begin{equation}
\hat{H} = P^2/(2m) + (m\omega^2/2) Q^2
\end{equation}
corresponding to the classical Hamiltonian
\begin{equation}
H(q,p) = p^2/(2m) + (mw^2/2) q^2.
\end{equation}
The density operator $\hat{\rho}$ for a {\em canonical} ensemble of such
oscillators at temperature $T$, with $\beta = 1/(kT)$, is given by the
relation
\begin{equation}
\hat{\rho} = [\exp (-\beta \hat{H})]/tr[\exp (-\beta \hat{H})].
\end{equation}
In this case the Wigner function is
\begin{eqnarray}
W(q,p) &=& [1/(\pi \hbar )] \tanh (\beta \hbar \omega /2) \exp \{
-[2/(\hbar \omega)] \nonumber\\
&&\times\tanh (\beta \hbar \omega /2)H(q,p)\}.
\end{eqnarray}
Inspection of (4.28) shows that in this case $W$ is well defined and
smooth as $\hbar \rightarrow 0$.  Indeed, in this limit $W$ becomes
the classical Boltzmann distribution (Liouville function),
\begin{equation}
\lim_{\hbar \rightarrow 0} W(q,p) = [\beta \omega /(2\pi)] \exp
[-\beta H(q,p)].
\end{equation}

However, in many cases the Wigner function becomes ever more highly
oscillatory as $\hbar \rightarrow 0$.  Such behavior is typical of
Wigner functions associated with ensembles that are relatively pure
\cite{7}.  For example, consider a perfectly pure ensemble consisting
of harmonic oscillators that are {\em all} in the {\em same} energy
eigenstate.  For such a pure ensemble, whose members are all in the
eigenstate $|n\rangle$ for which the energy has the value
\begin{equation}
E_n = (n + 1/2) \hbar \omega ,
\end{equation}
the Wigner function is given by the equation  \cite{5}
\begin{eqnarray}
W(n;q,p) &=& [(-1)^n/(\pi \hbar )] \exp [-2H(q,p)/(\hbar \omega )]
\nonumber\\ 
&&\times L_n^{(0)}[4H(q,p)/(\hbar \omega )].
\end{eqnarray}
Here $L_n^{(0)}$ is a Laguerre polynomial.

For simplicity of presentation, suppose units are selected in such a
way that both the spring constant $k$ and the mass $m$ have unit
value.  With this choice of units, the classical Hamiltonian becomes
\begin{equation}
H = (p^2 + q^2)/2,
\end{equation}
and $W$ takes the form
\begin{equation}
W(n;r) = [(-1)^n/(\pi \hbar )] \exp (-r^2/\hbar )L_n^{(0)} (2r^2/\hbar ).
\end{equation}
Here we have introduced a radius $r$ in phase space by the definition
\begin{equation}
r^2 = p^2 + q^2.
\end{equation}

Inspection of (4.33) shows that $W$ has an {\em essential} singularity
in $\hbar$ at $\hbar = 0$.  Moreover, differentiation operators, such
as $(\partial /\partial p)$, produce powers of $(1/\hbar )$.
Therefore, since (4.24) contains $(\partial /\partial \rho )^3$ and
higher derivatives, what appear to be quantum corrections actually
{\em diverge} as $\hbar\rightarrow 0$ like $(1/\hbar )$ and higher
powers of $(1/\hbar )$.  Direct evaluation of (4.33) gives the result
\begin{equation}
\lim_{\hbar \rightarrow 0} W(n;r) = 0 \ {\rm if} \ r > 0.
\end{equation}
Moreover, as can be checked directly, (3.12) also holds.  Therefore,
in the limit, $W(n;r)$ becomes a distribution that behaves like the
delta function distribution centered on the phase-space origin,
\begin{equation}
\lim_{\hbar \rightarrow 0} W(n,r) = [1/(2\pi r)] \delta (r) = (1/\pi )
\delta
(r^2) = \delta (p) \delta (q).
\end{equation}
Such a result is more or less to be expected since (4.30) shows that,
for {\em fixed} $n$, $E_n \rightarrow 0$ as $\hbar \rightarrow 0$.

A more interesting tack is to {\em increase} $n$ as $\hbar \rightarrow
0$ so that $E_n$ retains a constant value.  One might imagine that
then some classical features should emerge as $\hbar \rightarrow 0$.
Without loss of generality, the energy may be taken to have the value
1/2,
\begin{equation}
E = 1/2.
\end{equation}
So doing simply sets the scale in phase space.  From (4.30) and (4.37)
it follows that $\hbar$ is then given in terms of $n$ by the relation
\begin{equation}
\hbar = 1/(2n + 1).
\end{equation}
Correspondingly, the Wigner function takes the form
\begin{eqnarray}
W(n;r) &=& (-1)^n(1/\pi ) (2n + 1) \exp [-(2n + 1) r^2] \nonumber\\
&&\times L_n^{(0)} [(4n + 2)r^2].
\end{eqnarray}

Classically the Liouville density is concentrated on the unit circle
$r = 1$ when $E = 1/2$.  Figures 1 through 3 display $W(n;r)$ as a
function of $r$ for $n=10$, 20, and 40, respectively.  Evidently the
$\hbar\rightarrow 0$ $(n\rightarrow \infty )$ limit in this case is
far from simple, and requires some explanation.  However, two features
are immediately evident.  First, the Wigner function does decay
rapidly to zero for $r > 1$; but, contrary to classical expectations
based on the Liouville density, it does not decay for $r < 1$.
Instead it oscillates rapidly and even increases as $r$ decreases.
Indeed, since $L_n^{(0)}(0) = 1$, at the origin the Wigner function
has the value
\begin{equation}
W(n;0) = (-1)^n (1/\pi )(2n+1),
\end{equation}
which approaches $\pm \infty$ as $n \rightarrow \infty$.  Second, the
number of oscillations in the interval $[0,1]$ increases linearly with
$n$.  Consequently, derivatives such as $(\partial /\partial p)$
produce factors of $n$.  Equivalently, derivatives produce factors of
$(1/\hbar )$.  Therefore, as before, what appear to be quantum
correction terms in (4.24) actually {\em diverge} as $\hbar
\rightarrow 0$ like $(1/\hbar )$ and higher powers of $(1/\hbar )$.

\vspace{.4cm}
\centerline{\epsfig{figure=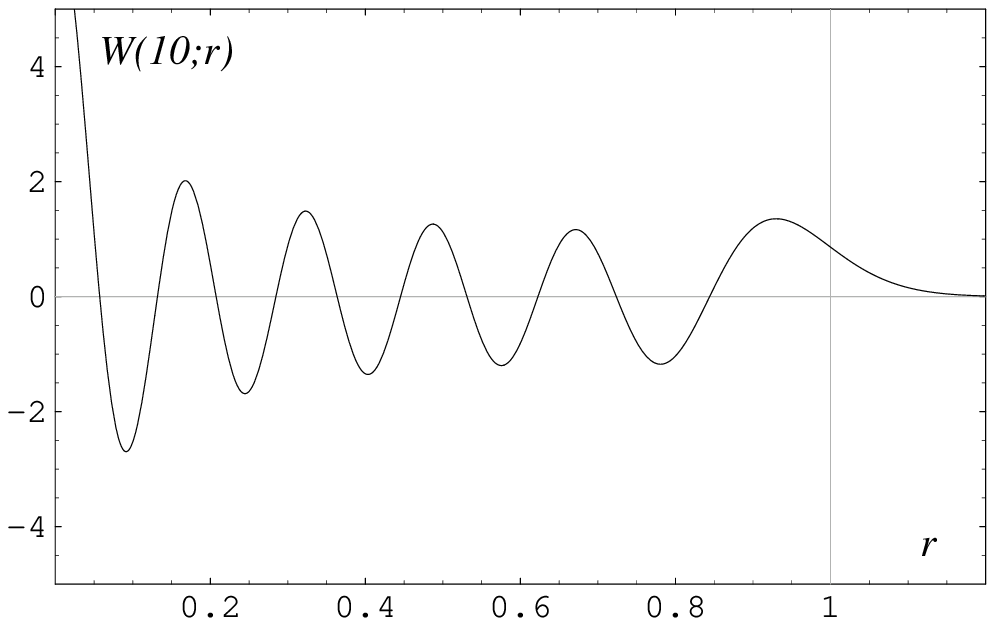,height=5cm,width=7cm,angle=0}}
\vspace{.35cm}
\centerline{{FIG. 1 {\small{The Wigner function for $n=10$.}}}}

\vspace{.4cm}
\centerline{\epsfig{figure=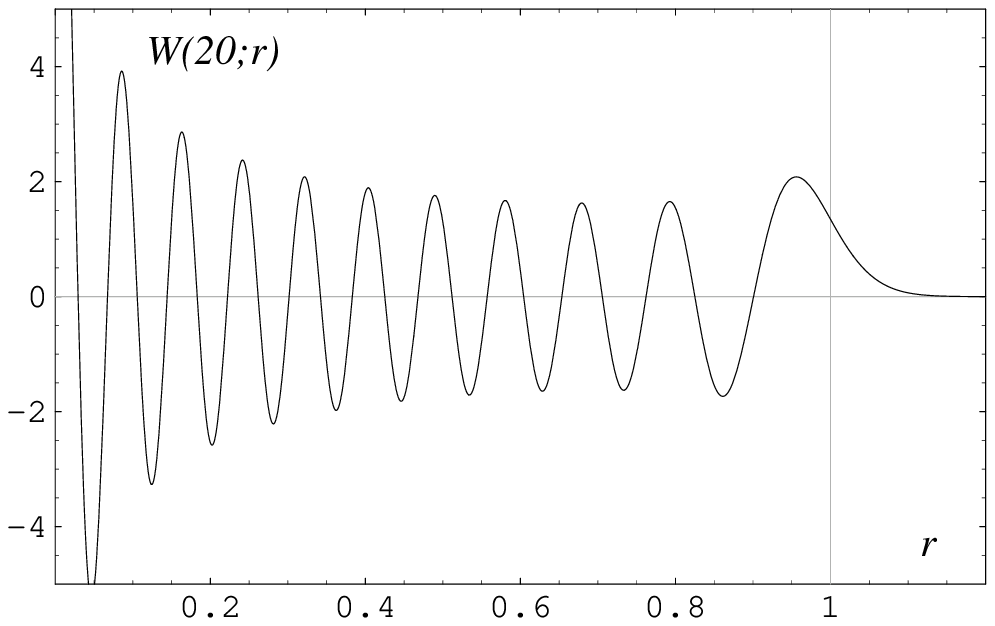,height=5cm,width=7cm,angle=0}}
\vspace{.35cm}
\centerline{{FIG. 2 {\small{The Wigner function for $n=20$.}}}}

\vspace{.4cm}
\centerline{\epsfig{figure=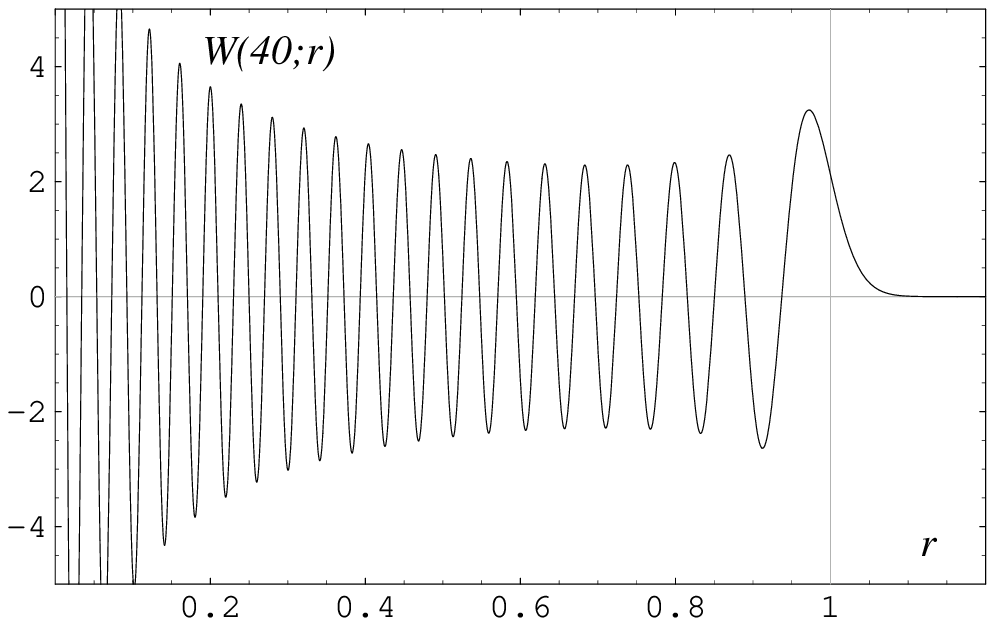,height=5cm,width=7cm,angle=0}}
\vspace{.35cm}
\centerline{{FIG. 3 {\small{The Wigner function for $n=40$.}}}}
\vspace{.4cm}

What sense can be made of the highly oscillatory behavior of $W(n;r)$
for $r< 1$?  One answer is again to view the Wigner function as a
distribution.  The density operator corresponding to an ensemble, all
of whose members are in the same energy eigenstate $|n\rangle$, is
given by the relation
\begin{equation}
\hat{\rho} = |n\rangle \langle n|.
\end{equation}
Apply (3.11) to the operator $(\hat{H})^{\ell}$ for $\ell =0$ and 1.
Making use of (3.3), (3.4), (4.25), (4.30), and (4.41) gives the
relation
\begin{eqnarray}
&&2\pi \int^{\infty}_0 dr r W(n;r) (r^2/2)^{\ell} = tr[(\hat{H})^{\ell}
|n\rangle \langle n|]\nonumber\\
&& = [(n+1/2) \hbar ]^{\ell}, \ \ell =0 \ {\rm and} \ 1.
\end{eqnarray}
Now put (4.38) into (4.42) to find the result
\begin{equation}
F(n,\ell ) = 2\pi \int^{\infty}_0 dr r^{2\ell + 1} W(n;r) = 1, \ \ell
=0 \ {\rm and} \ 1.
\end{equation}
It is tempting to conjecture that (4.43) holds for all positive
integer values of $\ell$.  However, some machinery beyond that already
presented is required for such a calculation because if $\hat{A}$ and
$\hat{B}$ are operators with Weyl correspondences $A$ and $B$, then it
is not generally the case that the Weyl correspondence of the product
$\hat{A} \hat{B}$ is the simple product $AB$ (even if $\hat{A}$ and
$\hat{B}$ commute, and even if they are equal).  Often there are
$\hbar$ dependent corrections.  They arise because there is an
ordering problem for mixed operator expressions of the form $Q^mP^n$.
The direct and inverse Weyl transforms (3.1) and (3.2), which were
used to define the Wigner function, imply a certain symmetric ordering
procedure \cite{4}.

By a suitable change of variable, (4.43) can be written in the form
\begin{equation}
F(n,\ell ) = (-1)^n [2(4n+2)^{\ell}]^{-1} G(n,\ell )
\end{equation}
where
\begin{equation}
G(n,\ell ) = \int^{\infty}_0 dx \exp (-x/2) x^{\ell} L_n^{(0)}(x).
\end{equation}
Like all classical polynomials, the Laguerre polynomials obey a 3-term
recursion relation.  In their case the recursion relation reads
\cite{6}
\begin{eqnarray}
&&xL^{(0)}_n(x) \nonumber\\
&&= -nL^{(0)}_{n-1}(x) + (2n+1)L^{(0)}_n(x) -
(n+1)L^{(0)}_{n+1}(x).\nonumber\\
\end{eqnarray}
It follows from (4.46) that the $G(n,\ell )$ obey the recursion
relation 
\begin{eqnarray}
&&G(n,\ell + 1) \nonumber\\
&&= -nG(n-1,\ell ) + (2n+1) G(n,\ell ) - (n+1) G(n+1,\ell
).\nonumber\\ 
\end{eqnarray}
Correspondingly, by (4.44), the $F(n,\ell )$ obey the recursion
relation 
\begin{eqnarray}
&&F(n,\ell + 1) \nonumber\\
&&= \left\{ \left[ n(4n-2)^{\ell}/(4n+2)^{\ell
+1}\right] F(n-1,\ell ) + (1/2) F(n,\ell ) \right.\nonumber \\
&&+ \left. \left[(n+1)(4n+6)^{\ell}/(4n+2)^{\ell +1}\right] F(n+1,\ell 
)\right\}.
\end{eqnarray}
Finally, repeated application of (4.48) yields for the first few
$F(n,\ell)$ the results  
\begin{eqnarray}
F(n,0) &=& 1, \nonumber\\
F(n,1) &=& 1, \nonumber\\
F(n,2) &=& 1 + [1/(2n+1)^2],\nonumber\\
F(n,3) &=& 1 + [5/(2n+1)^2],\nonumber\\
F(n,4) &=& 1 + [14/(2n+1)^2] + [9/(2n+1)^4)].
\end{eqnarray}
In view of (4.38), the terms in (4.49) involving $n$ are, in effect,
$\hbar$ dependent corrections.

Evidently (4.43) does not hold for $\ell$ larger than 1.  However
(4.49) shows that, for small $\ell$ and $n \rightarrow \infty$ ($\hbar
\rightarrow 0$), there is the limit
\begin{equation}
F(\infty , \ell )
= \lim_{n\rightarrow \infty} F(n,\ell ) = 1.
\end{equation}
Moreover, for large $n$ ($\hbar \rightarrow 0$) the recursion relation
(4.48) takes the form
\begin{eqnarray}
&&F(\infty , \ell + 1) \nonumber\\
&&= (1/4) F(\infty , \ell ) + (1/2) F(\infty , \ell )
+(1/4) F(\infty , \ell ) \nonumber \\
&&= F(\infty , \ell ).
\end{eqnarray}
Therefore (4.50) holds for all $\ell$.

Suppose the integral on the left side of (3.11) is to be calculated
for any $A(q,p)$ that is {\em polynomial} in the variables $q,p$, and
suppose that $W(q,p)$ depends only on the variable $r$.  Then the
integral may be factored into angular and radial parts, and a moment's
thought reveals that each angular integration vanishes unless the
accompaning power of $r$ is {\em even}.  Consequently, only integrals
of the form (4.43) occur in any such calculation.  Polynomials in
$q,p$ form a dense set of functions.  Therefore, in view of (4.43) and
(4.50), $W(n,r)$ has (in the distribution theoretic sense) the
$n\rightarrow \infty$ ($\hbar \rightarrow 0$) limit
\begin{equation}
\lim_{n \rightarrow \infty} W(n;r) \sim (1/2\pi ) \delta (r-1).
\end{equation}

Based on (4.52), one might conjecture that, in the limit $n
\rightarrow\infty$, most of the oscillations of $W(n;r)$, all of which
occur for $r < 1$, should {\em nearly average to zero} when integrated
over any smooth test function.  To examine this hypothesis, Figures 4
through 6 display the product $2\pi rW(n;r)$ as a function of $r$ for
$n=10$, 20, and 40, respectively.  Evidently there is a positive peak
near $r=1$ preceded by oscillations to the left.  The area under this
peak is found to be 1.2638, 1.2679, and 1.2704 for $n=10$, 20, and 40,
respectively.  If this peak along with the full oscillation preceding
it is considered, the areas are found to be 1.1387, 1.1460, and 1.1503
for $n=10$, 20, and 40, respectively.  Evidently the first few
oscillations do nearly ``saturate'' the relation
\begin{equation}
F(n,0) = \int_0^{\infty} dr 2\pi r W(n,r) = 1.
\end{equation}
However it is also evident that, even in the $n\rightarrow \infty$
limit, it is necessary to retain all oscillations to fulfill (4.53)
exactly.

\vspace{.4cm}
\centerline{\epsfig{figure=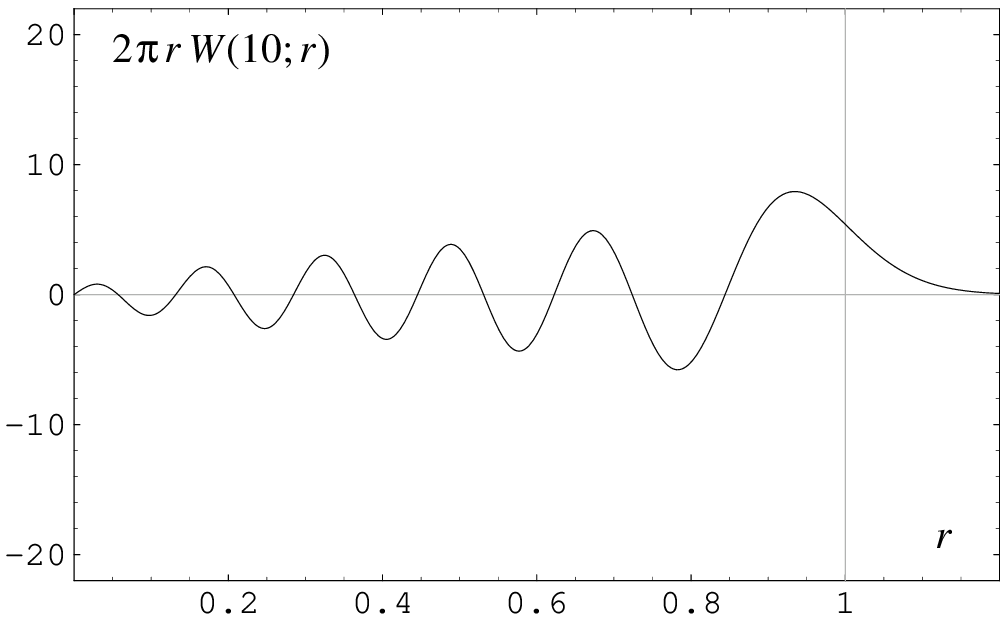,height=5cm,width=7cm,angle=0}}
\vspace{.35cm}
\centerline{{FIG. 4 {\small{The integrand in (4.53) for $n=10$.}}}}

\vspace{.4cm}
\centerline{\epsfig{figure=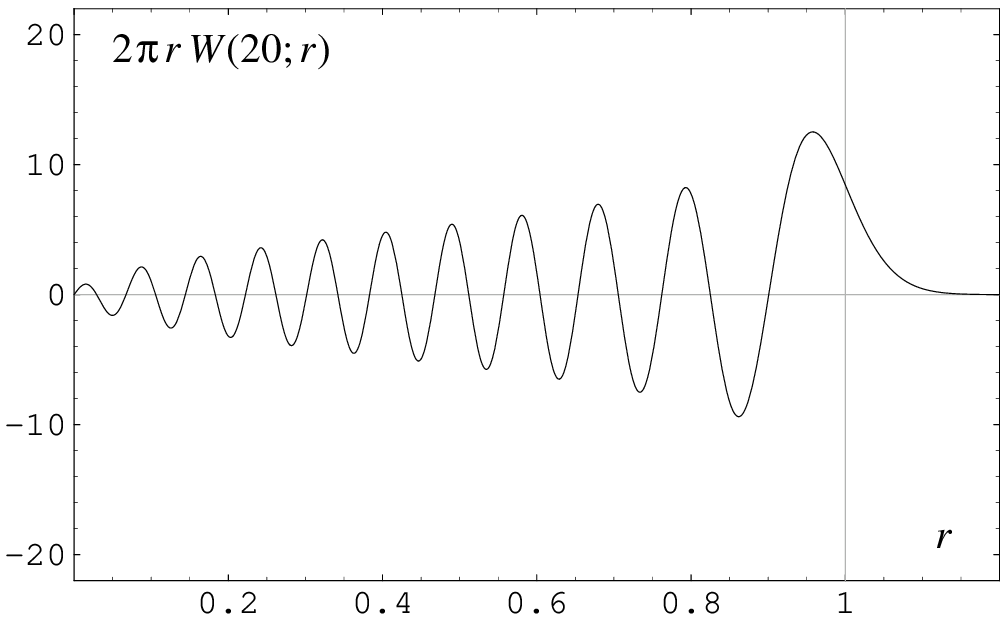,height=5cm,width=7cm,angle=0}}
\vspace{.35cm}
\centerline{{FIG. 5 {\small{The integrand in (4.53) for $n=20$.}}}}

\vspace{.4cm}
\centerline{\epsfig{figure=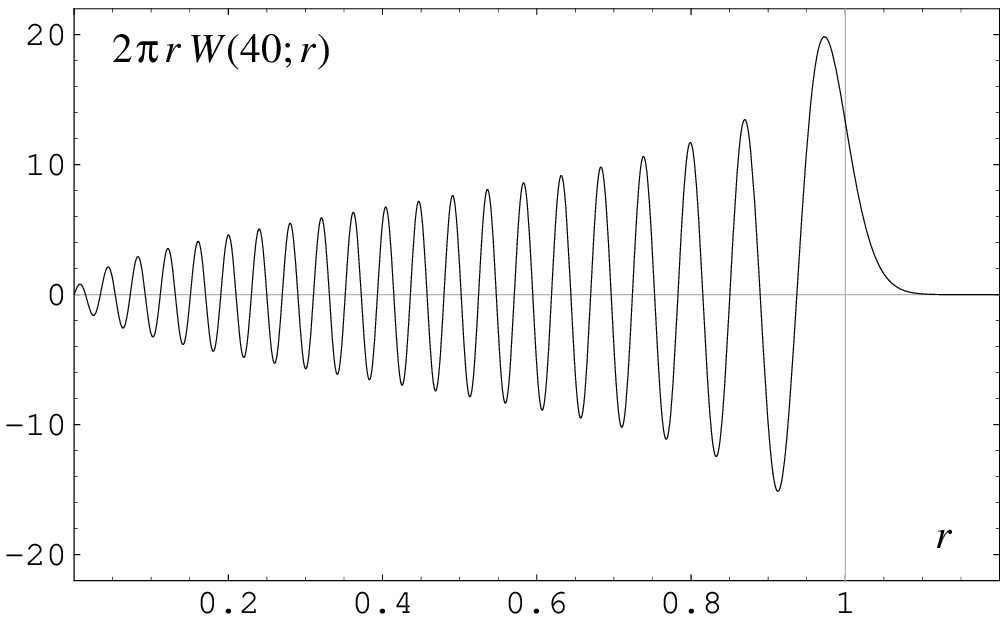,height=5cm,width=7cm,angle=0}} 
\vspace{.35cm}
\centerline{{FIG. 6 {\small{The integrand in (4.53) for $n=40$.}}}}
\vspace{.4cm}

\section{Discussion}
\label{sec:level5}
It has been shown that in general the Wigner function does not obey
(1.11) for nonlinear symplectic transformations, but rather there are
``quantum corrections'' as illustrated by (4.24).  This fact is no
surprise to those well acquainted with Wigner function literature
\cite{4}.  Let $H$ be a Hamiltonian of the form $T+V$ as in (3.4), and
assume that $T$ is quadratic in $p$.  Then it is known that the Wigner
function obeys the equation of motion \cite{4}
\begin{eqnarray}
\partial W/\partial t &=& -[W,H] \nonumber\\
&&- (\hbar^2/24)[(\partial /\partial q)^3H][(\partial /\partial p)^3W]
+ \cdots . 
\end{eqnarray}
Retaining only the first (Poisson-bracket) term on the right side of
(5.1) is equivalent to asserting (1.11).  Note also that if $H$
contains only linear and quadratic terms, which is equivalent to
considering only translation and linear symplectic maps, then the
correction terms in (5.1) vanish, and correspondingly (1.11) is exact.
Suppose $H$ is of the form
\begin{equation}
H = -(\alpha /3) q^3.
\end{equation}
The first correction term on the right side of (5.1) then gives the
result 
\begin{equation}
-(\hbar^2/24)(\partial /\partial q)^3 H = \alpha \hbar^2/12,
\end{equation}
in agreement with the first correction term in (4.24).

It is also well documented in the literature that Wigner functions are
often highly oscillatory, and that the $\hbar \rightarrow 0$ limit is
subtle \cite{8}.  Correspondingly, it is well recognized that the
correction terms in (5.1), just as the correction terms illustrated in
(4.24), must be treated with care \cite{9}.

\section{Acknowledgments}
Dr. Kwang-Je Kim first posed the question (1.11), and Dr. Dan T. Abell
carried out the numerical calculations associated with Figures 1
through 6.  This work was supported in part by U.S. Department of
Energy Grant DEFG0296ER40949.


\begin{references}
\bibitem[*]{byline} Electronic address: dragt@quark.umd.edu
\bibitem[\dagger]{byline} Electronic address: habib@lanl.gov
\bibitem{1} A.J. Dragt, Lie Methods for Nonlinear Dynamics with
Applications to
Accelerator Physics, University of Maryland Physics Department Report
(1998).
\bibitem{2a} E.P. Wigner, Phys. Rev. {\bf 40}, 749 (1932).
\bibitem{2} R.G. Littlejohn, Phys. Rep. {\bf 13}, 193 (1986).
\bibitem{3} A.J. Dragt, J. Opt. Sci Am. {\bf 72}, 372 (1982); Lie Algebraic
Methods for Ray and Wave Optics, University of Maryland Physics Department
Report (1998).
\bibitem{4} For a general discussion of Weyl transformations, Wigner
functions, and semiclassical approximations, see: \ M. Hillery, R.F.
O'Connell, M.O. Scully, and E.P. Wigner, Phys. Rep. {\bf 106}, 121 (1984);
V.I. Tatarskii, Usp. Fiz. Nauk. {\bf 139}, 587 (1983) [Sov. Phys. Usp. {\bf
26}, 311 (1983)]; V.F. Lazutkin, {\em KAM Theory and Semiclassical
Approximations
to Eigenfunctions}, Springer-Verlag (1993); A.M. Ozorio De Almeida, {\em
Hamltonian Systems: \ Chaos and Quantization}, Cambridge University Press
(1988); M.V. Berry and K.E. Mount, Rep. Prog. Phys. {\bf 35}, 315 (1972);
M.V.
Berry, ``Semiclassical Mechanics of Regular and Irregular Motion,'' in {\em
Chaotic Behavior in Deterministic Systems}, G. Iooss, R.H.G. Helleman, and
R.
Stora, eds., North Holland (1983); M.C. Gutzwiller, {\em Chaos in Classical and
Quantum Mechanics}, Springer-Verlag (1990); L.E. Reichl, {\em The
Transition to
Chaos in Conservative Classical Systems: \ Quantum Manifestations},
Springer-Verlag (1992).
\bibitem{5a} A.J. Dragt and D.T. Abell, ``Symplectic Maps and Computation
of
Orbits in Particle Accelerators,'' in {\em Integration Algorithms and
Classical Mechanics}, J.E. Marsden, G.W. Patrick, and W.F. Shadwick, eds.,
Fields Institute Communications {\bf 10}, American Mathematical Society
(Providence, RI, 1996); D.T. Abell, University of Maryland Physics
Department
Ph.D. thesis (1995).
\bibitem{7} For a discussion of important circumstances where there need
not be
a classical limit for the Wigner function describing close-to-pure states,
see O.
Penrose, Phil. Mag. {\bf 42}, 1373 (1951); O. Penrose and L. Onsager, Phys.
Rev. {\bf 104}, 576 (1956).  See also C.N. Yang, Rev. Mod. Phys. {\bf 34},
694
(1962).
\bibitem{5} H.J. Groenwold, Physica {\bf 12}, 405 (1946).
\bibitem{6} M. Abramowitz and I. Stegun, {\em Handbook of Mathematical
Functions}, National Bureau of Standards Applied Mathematics Series {\bf
55}
(1966).
\bibitem{8} M.V. Berry, Philos. Trans. Roy. Soc. {\bf 287}, 237 (1977);
M.V.
Berry and N.L. Balazs, J. Phys. A: \ Math. Gen {\bf 12}, 625 (1979);  E.J.
Heller, J. Chem. Phys. {\bf 67}, 3339 (1977); {\bf 65}, 1289 (1976).
\bibitem{9} S. Habib, Phys. Rev. D {\bf 42}, 2566 (1990).
\end{references}
\end{document}